\documentstyle[twoside,ijmpa,epsfig]{article}
\def\qed{\hbox{${\vcenter{\vbox{                        
   \hrule height 0.4pt\hbox{\vrule width 0.4pt height 6pt
   \kern5pt\vrule width 0.4pt}\hrule height 0.4pt}}}$}}

\newcommand{\bef}{\begin{figure}}
\newcommand{\eef}{\end{figure}}

\newcommand{\be}{\begin{equation}}
\newcommand{\ee}{\end{equation}}
\newcommand{\bea}{\begin{eqnarray}}
\newcommand{\eea}{\end{eqnarray}}
\newcommand {\mean}[1]{\left\langle #1 \right\rangle}

\def\bsc{{\sc a\kern-6.4pt\sc a\kern-6.4pt\sc a}}       
\def\bflatex{\bf L\kern-.30em\raise.3ex\hbox{\bsc}\kern-.14em
T\kern-.1667em\lower.7ex\hbox{E}\kern-.125em X}

\begin{document}
\runninghead{Using spectator distributions to measure the initial geometry
  fluctuation} {Md. Nasim}
\normalsize\textlineskip
\thispagestyle{empty}
\setcounter{page}{1}

\copyrightheading{}                     

\vspace*{0.88truein}

\fpage{1}
\centerline{\bf 
Using spectator distributions to measure the initial geometry
  fluctuation}
\vspace*{0.37truein}

\centerline{\footnotesize 
Md. Nasim \footnote{Corresponding author, e-mail: mnasim2008@gmail.com }}
\vspace*{0.015truein}
\centerline{\footnotesize Department of Physics $\&$ Astronomy,\\
University of California, Los Angeles, CA-90095, USA}

\vspace*{0.21truein}

\abstracts{
 A study of eccentricity ($\varepsilon_{n}$) fluctuations and its
possible impact on final state momentum anisotropy ($v_{n}$) for symmetric collisions are
presented in the framework of Glauber model. Effect of 
fluctuations of nucleon's  position on the initial geometry  has been
studied using a new method, where the difference between oppositely moving
spectators is taken as a measurement of eccentricity fluctuations.
This study shows that higher harmonics ($n$ =3, 4 and 5) of eccentricity
are less sensitive to  fluctuations in transverse plane  compared to
the 2$^{nd}$ harmonic. Position fluctuations in transverse plane will increase 
$\varepsilon_{2}$ and hence possibly $v_{2}$ for the most central nucleus-nucleus
collisions. For semi-central and peripheral collisions, the
fluctuations have opposite effect,  it deceases the eccentricity
$\varepsilon_{2}$.  The  fluctuation of initial geometry can be studied in
 collider experiments by studying the spectator distribution on the
both sides of the beam.
\\}{}{}

\vspace*{1pt}\textlineskip 
\vspace*{-0.5pt}

\noindent {PACS-key : 25.75.Ld, 25.75.-q} \\

\section{Introduction}
\label{sec:1}

One of the main goals of the high energy heavy-ion collision experiments is to study  the
QCD phase diagram~\cite{whitepapers,whitepapers1,whitepapers2,whitepapers3}. To achieve this goal, one has to
understand the properties of the system formed in such collisions.
 The momentum azimuthal angular anisotropy parameter $v_{n}$ has been
 considered as a good tool for studying
the system formed in the early stages of high energy collisions
at Super Proton Synchrotron (SPS), Relativistic Heavy Ion Collider
(RHIC) and Large Hadron Collider (LHC)~\cite{bes_res1,v2_BES_prc,v2_BES_prl,hydro,hydro1,hydro2,hydro3,hydro4}. It describes
the $n^{th}$ harmonic coefficient of the azimuthal Fourier decomposition of
the momentum distribution with respect to the reaction plane angle
($\Psi$)~\cite{method}. 
The final state momentum anisotropy ($v_{n}$) reflects  the hydrodynamic response of
initial spatial anisotropy ($\varepsilon_{n}$). 
According to hydrodynamical description, $v_{n}$ is sensitive to the
geometry of initial state of the system formed in the
collision as well as the hydrodynamic evolution governed by the
equation of state of the matter~\cite{hydro,hydro1,hydro2,hydro3,hydro4,early_v2}.\\
Knowing the initial geometry and fluctuations in heavy-ion
collisions has recently been shown to have important consequences on
interpreting the experimental data from various experiment at RHIC
and LHC. Experimentally measured non zero  odd harmonic ($n$$\geq$3)  has been
interpreted  as the result of statistical fluctuations in the transverse
positions (according to uncertainty principle)
of nucleons undergoing hadronic scattering. Moreover, measured $v_{2}$
cannot be described by an smooth initial energy density
distribution, unless
one includes flow fluctuations arising due to the eccentricity
fluctuations in the calculations~\cite{v2flucth1,v2flucth2}. In addition to $v_{n}$, there are other
experimental observables which cannot be explained without including
eccentricity fluctuations. For example, dihadron
correlations in azimuthal angle~\cite{mach} and
pseudorapidity~\cite{ridge,ridge1}. The contribution from the odd harmonics associated with the particle azimuthal angle
distribution to dihadron correlations is found to be an important
factor. \\
Several phenomenological studies have been  carried out on initial geometry
anisotropy and fluctuations to understand its influence on
experimental data~\cite{Bmuller,Jwang,TRenk,Lxhan,alver,rihan}. The aim of this paper is to discuss the centrality dependence of
various harmonics of 
initial spatial anisotropy and its sensitivity to the fluctuation in
position of nucleons using a new method. In this paper, $\varepsilon_{n}$ are calculated within
a framework of Monte Carlo Glauber (MCG) model~\cite{gbl},  which allows the
generation of collisions with event-by-event-fluctuating initial condition. 
The paper is organized in the following way. In section \textrm{2},  Glauber
model has been briefly discussed. Section \textrm{3} describes the
study of $\varepsilon_{n}$ and its sensitivity to the fluctuations
using the MCG model. Finally, the summary has been given in section \textrm{4}.

\section{Model Description}
\label{sec2}
In MCG model, the nuclear distribution function inside a
nucleus is taken to be of the Woods-Saxon form,
\begin{equation}
\rho_{A}(r)=\frac{\rho_{0}}{1+exp[(r-R)/d]},
\end{equation}
where the radius ($R$) and the diffuse constant ($d$) are taken as
$R$ = 6.38 fm, $d$ = 0.535 fm for  Au nucleus. In this model,
nuclei are assembled by positioning the nucleons randomly in
a three-dimensional coordinate system, on an event-by-event basis, according
to the  Woods-Saxon density profile. A collision between two nuclei  is
considered as a sequence of independent nucleon-nucleon collisions. 
 In a nucleus-nucleus collision, two nucleons with transverse distance
 $d$ $\leq$ $\sqrt{\sigma_{NN}/\pi}$ 
will collide with each other.
Here $\sigma_{NN}$ is the total nucleon-nucleon cross-section.

Using the transverse position
coordinates of each colliding nucleon, various moments of participant
eccentricity~\cite{alver} have been calculated as:
\begin{equation}
  \varepsilon_{n} = \frac{\sqrt{\mean{r^2\cos(n\varphi_{part})}^2 + \mean{r^2\sin(n\varphi_{part})}^2}}{\mean{r^2}},
\label{eq:ecc}
\end{equation}
where $r$ and $\varphi_{part}$ are the polar coordinate positions of participating nucleons.
\begin{equation}
  r=\sqrt{x^{2}+y^{2}} 
\label{eq:ecc2}
\end{equation}
and
\begin{equation}
 \varphi_{part} =tan^{-1}(y/x).
\label{eq:ecc3}
\end{equation} 
\begin{figure}
\centerline{\includegraphics[scale=0.35]{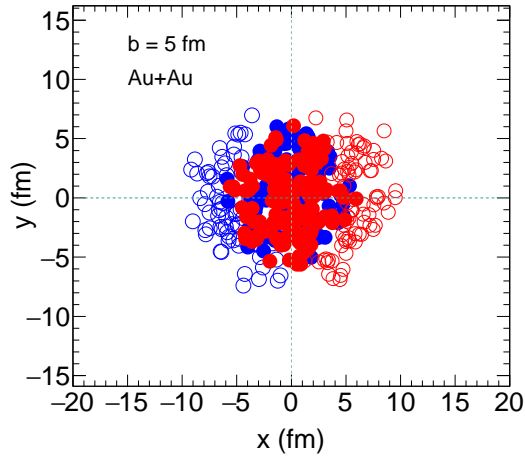}}
\caption{(Color online) Distribution of nucleons in the transverse plane for one typical
Au+Au event at $b$ = 5 $fm$. Solid and open circle represents the
participants and spectators nucleons, respectively, from two
colliding nuclei.}
\label{n_xy}
\end{figure} 
In this study,  approximately 7 million events for each configuration with fixed impact
parameter $b$ = 1 $fm$, 5 $fm$, and  9 $fm$, are
generated for Au+Au collisions with $\sigma_{NN}$ = 40 $mb$. Variation in
$\sigma_{NN}$ does not change results qualitatively. 
A standard MCG model code which is used as an input in AMPT model~\cite{ampt,ampt1} has
been used to generate events in this study.
The distribution of  nucleons in the transverse plane for a single
 event is shown in Fig.~\ref{n_xy}.

\section{Results and Discussion}
\label{sec3}
\begin{figure}
\centerline{\includegraphics[scale=0.4]{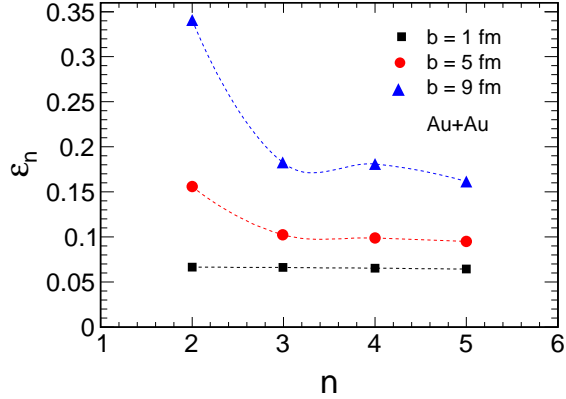}}
\caption{(Color online) Magnitude of spatial eccentricity in transverse plane for different
  harmonics in Au+Au collisions at $b$ = 1
  $fm$, $b$ = 5 $fm$ and $b$ = 9 $fm$. }
\label{en_n}
\end{figure}

\begin{figure}
\centerline{\includegraphics[scale=0.4]{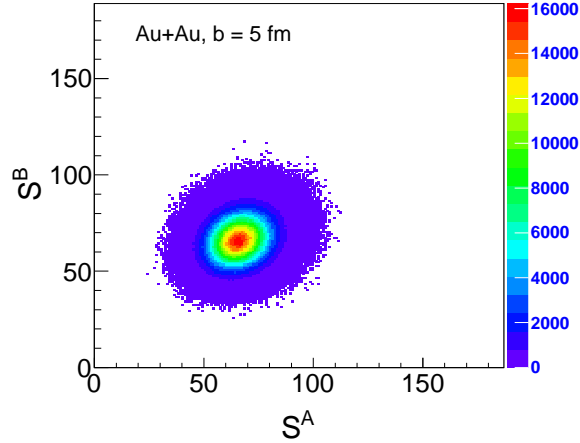}}
\caption{(Color online) Number of spectators in A-nucleus ($S^{A}$) vs
  Number of spectators in B-nucleus ($S^{B}$) in Au+Au collisions at  $b$ = 5 $fm$. }
\label{sa_sb}
\end{figure}

\begin{figure}
\centerline{\includegraphics[scale=0.4]{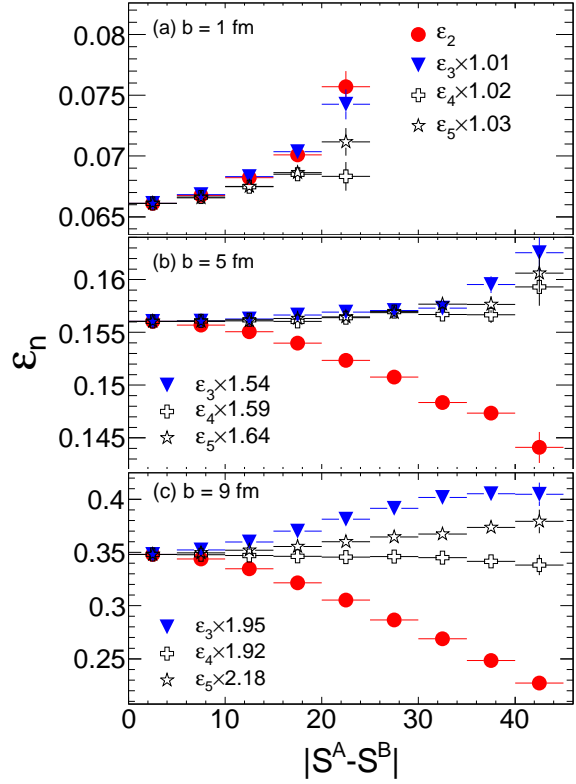}}
\caption{(Color online) Spatial eccentricity $\varepsilon_{2}$,
  $\varepsilon_{3}$,  $\varepsilon_{4}$,  and $\varepsilon_{5}$
  versus $|S^{A}-S^{B}|$ for collisions at   (a) $b$ = 1 $fm$,  (b) $b$
  = 5  $fm$ and (c) $b$ = 9  $fm$}
\label{en_sab}
\end{figure}

\begin{figure}
\centerline{\includegraphics[scale=0.5]{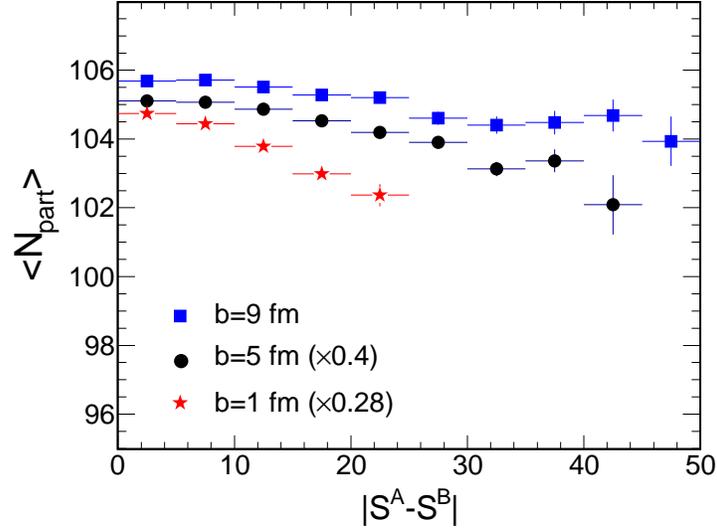}}
\caption{(Color online) Average $N_{part}$ versus $|S^{A}-S^{B}|$ for collisions at  $b$ = 1 $fm$,  $b$
  = 5  $fm$ and $b$ = 9  $fm$. Here $<N_{part}>$ for collisions at
  $b$ = 1 $fm$ and  5 $fm$ are scaled by factor 0.28 and 0.4, respectively. }
\label{npart_sab}
\end{figure}

Fig.~\ref{en_n} shows magnitude of spatial initial eccentricity in transverse plane for different
harmonics (from $n$=2 to $n$=5) in Au+Au collisions  at $b$ =
1 $fm$, $b$ = 5 $fm$ and $b$ = 9 $fm$ . The large value of
$\varepsilon_{2}$ for  $b$ = 5 $fm$ and $b$ = 9 $fm$ is due to initial elliptic shape
of the overlapping region in a collision of large impact
parameter. All odd higher harmonics ($n$ $>$ 2) are  generated
due to fluctuations in  transverse positions of nucleons. For a
nucleus with  smooth density distribution, all odd higher harmonics ($n$
$>$ 2) of $\varepsilon_{n}$ will be zero. In case of nucleus-nucleus
collisions at $b$ = 1 $fm$, the initial overlapping geometry is almost
isotropic, hence magnitude of $\varepsilon_{2}$ is small and
comparable with values of  $\varepsilon_{3}$, $\varepsilon_{4}$ and
$\varepsilon_{5}$. Centrality dependence of  $\varepsilon_{2}$ can be understood, since it reflects the anisotropy of overlapping region of
two nuclei. But we observed, as shown in Fig.~\ref{en_n}, that all
higher harmonics of $\varepsilon_{n}$ reveal similar
centrality dependence, like to that $\varepsilon_{2}$. The
fluctuations behave differently for collisions with different impact parameter.
Although the non-zero $\varepsilon_{2}$
is originated due to initial elliptic shape, it can be modified by the
nucleon density fluctuations. Therefore, it is very important to
understand the role of initial geometry fluctuations in heavy-ion collisions.
The main purpose of this paper is to investigate how various harmonics of eccentricity
change with nucleon density fluctuations. \\
The fluctuation varies on event-by-event basis. Therefore, the number of
spectators ($S$) and participants ($N_{part}$) will also vary on event-by-event
basis. Moreover, in a single event,  number of spectators from target nucleus (labeled
as $A$) and projectile nucleus (labeled as $B$) can be different due
to the fluctuation of nucleon's position in the transverse
plane~\cite{sandeep}. 
In this study, the difference in the number of spectator  between two colliding
nuclei ($|S^{A}-S^{B}|$) has been used to quantify the fluctuation, more fluctuation
means large difference  and vice-versa.
Total number of events for a fixed impact parameter are divided in
several sub-groups based on $|S^{A}-S^{B}|$.
\\
Fig.~\ref{sa_sb} shows, event-by-event distribution of number of
spectators in A-nucleus ($S^{A}$) and in B-nucleus($S^{B}$) from MCG
model at $b$ = 5 $fm$. The maximum difference
between $S^{A}$ and $S^{B}$ can be of the order of 50.\\
Values of  $\varepsilon_{2}$, $\varepsilon_{3}$, $\varepsilon_{4}$,
and $\varepsilon_{5}$ as function of $|S^{A}-S^{B}|$
are shown in Fig.~\ref{en_sab}. Panel (a), (b) and (c) corresponds to
events with fixed $b$ =1, 5 and 9 $fm$, respectively. In each  case, 
magnitude $\varepsilon_{3}$, $\varepsilon_{4}$,
and $\varepsilon_{5}$ are scaled to match their values with
$\varepsilon_{2}$ at $|S^{A}-S^{B}|$ =5. For central events with  $b$ =1 $fm$ (i.e. panel(a)),  all harmonics
of $\varepsilon_{n}$ increases with increase in $|S^{A}-S^{B}|$. This
indicates that, fluctuations enhance the anisotropy for the most central
collisions. As we have fixed the value of impact parameter,
changes in $\varepsilon_{n}$ is entirely due to  fluctuations. 
What is striking in this observation is that, 
fluctuations making the system more elliptic, while also
makes the system more triangular, quadratic and pentagonal. \\
Now for  semi-central events with $b$ =5 $fm$ (panel(b)),  we can see
that there is small change in eccentricity for $n$ $\geq$ 3, their values are increasing with increase in $|S^{A}-S^{B}|$. 
But we observed that the magnitude of  $\varepsilon_{2}$ decreases with increase
in $|S^{A}-S^{B}|$, unlike central events. This shows that the fluctuations in the transverse
plane are decreasing initial elliptical geometry a nucleus-nucleus
collision.\\
For peripheral events with  $b$ = 9 $fm$ (panel(c)),
$\varepsilon_{2}$ changes sharply with change in $|S^{A}-S^{B}|$, and
shows a decreasing  trend with increasing fluctuations, like semi-central events.
The $\varepsilon_{3}$ and $\varepsilon_{5}$ increases with
fluctuations and almost negligible change for $\varepsilon_{4}$ with
respect to other harmonics. \\
To quantify the sensitivity of $\varepsilon_{n}$ to fluctuations,
ratios between maximum ($\varepsilon_{n}^{max}$) and minimum
($\varepsilon_{n}^{min}$) value of
eccentricity has been  calculated. The maximum change in
$\varepsilon_{2}$, $\varepsilon_{3}$,
$\varepsilon_{4}$ and $\varepsilon_{5}$  are $\sim$ 15$\%$, 12$\%$,
3$\%$ $\&$ 7.5$\%$;  $\sim$ 8.3$\%$, 4.5$\%$, 2$\%$ $\&$ 3.5$\%$;
and   $\sim$ 52$\%$, 14$\%$, 3$\%$ $\&$ 8.5$\%$; for $b$ =1, 5 and 9
$fm$, respectively. This indicates that of all the harmonics,
$\varepsilon_{2}$ is more sensitive to the fluctuation and then
$\varepsilon_{3}$. This observation is consistent with the
previous study done using different asymmetric collision in AMPT
model~\cite{rihan}, where it was shown that $v_{2}$ is more sensitive
than  $v_{3}$.
On the other hand,  we observed that the
$\varepsilon_{4}$ is less sensitive to the fluctuations compared to
$\varepsilon_{5}$.\\
We know that the final state momentum anisotropies  are driven by the initial
spatial anisotropy and flow coefficients ($v_{n}$) are proportional to
$\varepsilon_{n}$. Therefore the change in $\varepsilon_{n}$ due to
fluctuation will affect $v_{n}$ in similar manner. Sensitivity of the $v_{n}$ to
the fluctuation can be different compared to $\varepsilon_{n}$ and
that depends how the $\varepsilon_{n}$  evolve through different stages of the fireball history and translate
into final-particle momentum anisotropies. But qualitatively,
sensitivity of the $v_{n}$ could  be similar  like
$\varepsilon_{n}$. Therefore, from  Fig.~\ref{en_sab} we expect that
the initial fluctuations in transverse plane  will generate more $v_{2}$ in  most central collisions,
whereas for semi-central to peripheral collision magnitude of $v_{2}$
will be reduced due to the fluctuations. This observation is in
agreement  with previous study carried out using  3+1 Viscous
Hydrodynamics in Ref~\cite{v2flucth2}. \\
In real collider experiment, one can easily measure number of spectator in both direction of beam and
hence the difference between them. Therefore, using this method one can
identify the events with different amount of fluctuations in one
centrality bin and can measure $v_{n}$ to understand the effect of
fluctuation. Narrow centrality bins will be more appropriate for
this study.  In experiment, centrality is usually estimated by
number of particles produced in the collision. The
number of spectators is anti-correlated with the number of particle
participating nucleons in the collisions. We have studied
how  the average $N_{part}$ changes with $|S^{A}-S^{B}|$, as shown in
Fig.~\ref{npart_sab}, to estimate the effect of event mixing from
different centrality group. We can see from Fig.~\ref{npart_sab} that the
change in $<N_{part}>$ is less than 3$\%$.\\

\section{Summary}
\label{sec:9}
A study on initial collision geometry fluctuations for a symmetric
system using MCG model has been presented. 
It has been observed that all other higher
harmonics of $\varepsilon_{n}$ show centrality dependence  like $\varepsilon_{2}$.
A new method using number of spectator nucleons has been used to
separate events with different amount of fluctuations. Due to
 fluctuations in the transverse plane of colliding nuclei,
 $\varepsilon_{n}$ (and possibly $v_{n}$) increases in most central collision. For semi-central and
peripheral collisions,  $\varepsilon_{2}$ (and possibly $v_{2}$) is minimised by the fluctuations,
on the other hand all other higher harmonics are found to be higher due  the
fluctuations in the transverse plane. Moreover, we observed that 2$^{nd}$
harmonic is more sensitive to the collision geometry fluctuation
compared to higher harmonics although higher harmonics are generated
due to fluctuations. This new proposed method can be applied to more
realistic transport model  and in real experiment to study the
fluctuations in  $v_{n}$, which is very crucial
to understand various properties like transport coefficient of the 
system created in heavy-ion collisions.\\
Only the fluctuation of nucleon's position in the transverse
plane has been discussed in this paper. For ultra-relativistic nucleus-nucleus
collisions, nuclei are contracted along beam axis (Z-axis) and looks
like as thin plates in lab-frame. As a result, fluctuations of
nucleons position along longitudinal directions are negligibly
small. But in case of collisions at a low energy, like AGS energy,
where the  out-of-plane squeeze-out phenomena in elliptic flow was observed, fluctuations
along  longitudinal directions may not be negligibly small. 
Future investigation can be done in this direction using various transport model to see
the effect of longitudinal fluctuations in the final states momentum anisotropies
using this new proposed methods.

\noindent {\bf Acknowledgments : }{
This work is supported by
the DOE Grant of Department of Physics and Astronomy, UCLA, USA.
}

\end{document}